\documentclass[12pt]{article}
\usepackage{amsfonts}
\usepackage{amssymb}
\usepackage{amsmath}
\usepackage{graphicx}
\usepackage{prettyref}
\usepackage{slashed}
\usepackage{setspace}
\usepackage{verbatim}
\usepackage{enumerate}

\usepackage{color}

\definecolor{darkgreen}{rgb}{0,0.5,0}
\definecolor{darkblue}{rgb}{0,0,0.6}
\definecolor{purple}{rgb}{0.4,.2,0.7}
\usepackage[colorlinks=true,citecolor=darkgreen,linkcolor=purple,urlcolor=purple]{hyperref}

\newcommand{\cn}{{\cal N}}

\newcommand{\reef}[1]{(\ref{#1})}

\newcommand{\cl}{{\cal L}}

\newcommand{\be}{\begin{equation}}
\newcommand{\ee}{\end{equation}}

\def\be{\begin{equation}}
\def\ee{\end{equation}}
\def\bea{\begin{eqnarray}}
\def\eea{\end{eqnarray}}
\def\ba{\begin{array}}
\def\ea{\end{array}}
\def\bd{\begin{displaymath}}
\def\ed{\end{displaymath}}

\def\Tr{{\rm Tr}}

\def\a{\alpha}
\def\b{\beta}

\def\d{ \delta}
\def\e{\epsilon}           
  
\def\g{\gamma}
\def\h{\eta}

\def\k{\kappa}                    
\def\l{\lambda}
\def\m{\mu}
\def\n{\nu}
\def\o{\omega}  
\def\r{\rho}                                     
\def\s{\sigma}                                   

\def\D{ \Delta}

\def\pa{\partial}                              
\def\>{\rangle} 
\def\<{\langle} 
\def\Dsl{D \hskip-.6em \raise1pt\hbox{$ / $ } }
\def\to{\rightarrow}
\def\pa{\partial}
\def\lab{\label}

\newcommand{\eps}{\epsilon}

\begin{document}
\numberwithin{equation}{section}

\begin{titlepage}

 \begin{flushright}
{\tt MIT-CTP-4538}
\end{flushright}

\vspace*{2.3cm}

\begin{center}
{\LARGE \bf de Sitter Supersymmetry Revisited} \\

\vskip15mm

{\bf Tarek Anous$^{1}$,  Daniel Z. Freedman$^{1,2,3}$ and Alexander Maloney$^{4,5}$}
\vskip5mm

{\it $^{1}$Center for Theoretical Physics, Massachusetts Institute of Technology}

{\it $^{2}$Department of Mathematics, Massachusetts Institute of Technology}

{\it $^{3}$Stanford Institute for Theoretical Physics, Department of Physics, Stanford University}
 
{\it $^{4}$Departments of Physics, McGill University} 
 
{\it $^5$Center for the Fundamental Laws of Nature, Harvard University}

\vskip5mm
{\tt tanous@mit.edu, dzf@math.mit.edu, maloney@physics.mcgill.ca}  \\

\vskip15mm

\begin{abstract}  
We present the basic $\cn =1$ superconformal field theories in
four-dimensional de Sitter space-time,  namely the non-abelian
super Yang-Mills theory and the chiral multiplet theory with gauge interactions or cubic superpotential. 
These theories have eight supercharges and are invariant under the full $SO(4,2)$ group of conformal symmetries, which includes the de Sitter isometry group $SO(4,1)$ as a subgroup.
The theories are ghost-free and the anti-commutator $\sum_\a\{Q_\a, Q^{\a\dagger}\}$ is positive. SUSY Ward identities uniquely select the Bunch-Davies vacuum state.
This vacuum state is invariant under superconformal transformations,
 despite the fact that de Sitter space has non-zero Hawking temperature.
The $\cn=1$ theories are classically invariant under the $SU(2,2|1)$ superconformal group, but this symmetry is broken by radiative corrections. 
However, no such difficulty is expected in the $\cn=4$ theory, which is presented in appendix \ref{n4ap}. 

%


\end{abstract}
\end{center}

\end{titlepage}
\section{Introduction}
In four dimensional de Sitter space (dS$_4$) there are no known $\cn=1$ unitary, supersymmetric theories, while such theories abound in Minkowski and anti-de Sitter space.
The two main obstacles are (see \cite{Pilch:1984aw,Lukierski:1984it}):
\begin{itemize}
\item
Majorana Killing spinors, which are generally necessary for the construction of $\cn=1$ Lorentzian signature SUSY theories in curved space, do not exist in dS$_4$.
 \item
The usual de Sitter super-algebra -- in which the $\{Q,\bar{Q}\}$ anti-commutator closes on the generators of the $SO(4,1)$ de Sitter isometry group -- has no unitary representations.  
Indeed, one can show that $\sum_\a~\{Q_\a,Q^{\a\dagger}\}=0$, so any non-trivial representation of the de Sitter superalgebra has negative norm states \cite{Pilch:1984aw,Lukierski:1984it}.
\end{itemize}
In this note we consider a modest exception to the above no-go results: global superconformal theories in dS$_4$.
These theories are consistent  because 
dS$_4$ is locally conformal to Minkowski space.  Indeed, the actions and transformation rules are determined by an appropriate Weyl transformation 
of the flat space theory.  

Our primary focus is the basic classical  $\cn=1$ superconformal theories in dS$_4$.  We avoid the first obstacle noted above by constructing our theories using conformal Killing spinors, for which there is no difficulty in imposing a Majorana condition.  
From these one obtains 8 
conserved supercharges whose anti-commutators 
give the 10 $SO(4,1)$ Killing charges along with 5 conformal Killing charges 
necessarily to fill out the full $SO(4,2)$ conformal algebra of de Sitter space.
Thus the second obstacle noted above is avoided because the $\{Q,\bar{Q} \}$ algebra closes on $SO(4,2)$, rather than $SO(4,1)$,  and 
\be \lab{poscom}
\sum_\a\{Q_{\a}, Q^{\dagger\a}\}  = Q[K]  \ge 0\,,
\ee
in which $Q[K]$ is a formally conserved conformal Killing charge  with positive integrand when expressed as an integral over a time slice.


There is interesting previous work.
First, the positive energy theorem for  de Sitter space proven by Kastor and Traschen \cite{Kastor:2002fu} (see also \cite{Shiromizu:2001bg}) exploits, as we also do, the fact that the  coset $SO(4,2)/SO(4,1)$  contains a conformal energy (i.e. a generator of conformal time translations) with positive spectrum.   Second,  
  de Medeiros and Hollands \cite{deMedeiros:2012sb, deMedeiros:2013jja, deMedeiros:2013mca}  have studied $\cn=2$ superconformal supersymmetry in conformally flat spacetimes such as dS$_4$.
It is puzzling that they state that  
 $\cn=1$ superconformal theories cannot be constructed in dS$_4$. 
The geometries that support  conformal Killing spinors are studied in \cite{ deMedeiros:2013mca, Cassani:2013dba}.
Several authors have also considered theories with rigid supersymmetry in Euclidean curved spaces \cite{Festuccia:2011ws, Dumitrescu:2012ha} and superconformal symmetry in Lorentzian curved spacetimes\cite{Cassani:2012ri, Klare:2013dka}.\footnote{Recent work \cite{Pahlavan:2005pm,  Masouminia:2013oza} also develops de Sitter symmetry from superfields by embedding dS$_4$ in flat space.  
As the super-algebra closes on $SO(4,1)$, we suspect this theory is non-unitarity.}

Recently, in light of the dS/CFT correspondence \cite{Witten:2001kn,Strominger:2001pn,Maldacena:2002vr}, and its first realization in terms of a higher spin theory in dS$_4$ \cite{Anninos:2011ui}, some progress has been made in writing down supersymmetric higher spin theories in dS$_4$ \cite{  Vasiliev:1986qx,Sezgin:2012ag}. However, the known examples have $\mathcal{N}=2$ supersymmetry.

In the following sections we will construct the non-abelian
super Yang-Mills and the chiral multiplet theories in de Sitter space.
We will discuss
why the $\mathcal{N}=1$ superconformal field theories on dS$_4$ satisfy \reef{poscom}. We also explore the SUSY Ward identities that relate propagators of boson and fermion fields.
Our $\cn  = 1$ theories are conformal invariant only at the classical level, since perturbative radiative corrections introduce a scale and an RG flow. We also present $\cn=4$ SYM on dS$_4$ in appendix \ref{n4ap}, which we expect enjoys exact $SU(2,2|4)$  supersymmetry.

\section{De Sitter supersymmetry and conformal Killing spinors}

In a general supersymmetric field theory, the commutator of two SUSY transformations with spinor parameters $\e',~ \e$ has the structure (on any field $\Phi(x)$):
\be\lab{salg0}
[\d_{\e'}\,,\,\d_\e] \Phi=- (\bar\e'\g^\m\,\e) D_\m\Phi + \ldots\,,
\ee 
where $K^\m =  \bar\e'\g^\m\,\e $ is usually a Killing vector.  With this perspective in view let us compare the situation in AdS$_4$ and dS$_4$.\footnote{We present our Dirac matrix conventions in appendix \ref{conventions}.}

\paragraph{SUSY in AdS$_4$:}   
Suppose that $Y^A,~ A=0,1,2,3,4$ are Cartesian coordinates in a 5-dimensional space with metric $\eta_{AB} = {\rm diag} (-+++-). $ Then AdS$_4$ can be viewed as  the embedded hyperboloid satisfying $
\h_{AB}Y^AY^B=- a^2\,.$  The usual Killing spinor condition and its Dirac adjoint
\be  \lab{adsksp}
D_\mu \e =  \frac{1}{2a}\g_\m \e    \qquad\qquad   \bar\e \overleftarrow{D}_\m = -  \frac{1}{2a}\bar\e \g_\m
\ee
are compatible with the integrability condition $[D_\m,D_\n]\e = \tfrac14 R_{\m\n ab}\g^{ab} \e$ in a constant negative curvature spacetime.  They are also real (in a Majorana representation) and admit a maximal set of Majorana solutions.  Further, the bilinear $K^\m =  \bar\e'\g^\m\,\e $ satisfies
\be \lab{kvec}
D_\m K_\n +D_\n K_\m =0\,.
\ee
It is a Killing vector, and the set of such bilinears spans the set  of 10 Killing vectors of $SO(3,2)$.   This leads to the well known and understood theory of AdS SUSY.   

\paragraph{SUSY in dS$_4$?}   The embedded hyperboloid of dS$_4$ is given by  $\h_{AB}Y^AY^B= a^2\,$  where $Y^A$ are now coordinates in a 5-dimensional space with metric  $\eta_{AB} = {\rm diag} (-++++). $
The would-be Killing spinor condition\footnote{The alternate condition in which $i\g_\m \to \g_5\g_\m $ on the right side differs by a chiral rotation of the spinors and is thus equivalent \cite{Kastor:2002fu}.  } 
\be \lab{dsksp}
D_\mu \e =  \frac{i}{2a}\g_\m\e    \qquad\qquad   \bar\e \overleftarrow{D}_\m =  \frac{i}{2a}\bar\e \g_\m
\ee
is required by integrability in dS$_4$,  which has constant positive curvature.
There is a maximal set of solutions. However,  the Killing spinor equation is essentially complex, so  there are no Majorana spinor solutions.  Further,  the spinor bilinears $K^\m =  \bar\e'\g^\m\,\e $ satisfy
\be\lab{ckvec}
D_\m K_\n +D_\n K_\m = \tfrac12 g_{\m\n}D_\r K^\r =  \frac{i}{a} g_{\m\n}(\bar\e'\e)\,.
\ee 
These bilinears are conformal Killing vectors, since the right hand side is proportional to $g_{\mu\nu}$; the ``weights" (the coefficients of $2 g_{\mu\nu}$) are
$\tfrac14 D_\r K^\r$. 

This is one clue that superconformal  SUSY is needed in de Sitter space.  However  spinors that satisfy \reef{dsksp} cannot be used, because their bilinears are complex, and their appearance in \reef{salg0} is not consistent with reality properties of the fields.

This problem can be  bypassed by working with conformal Killing spinors (CKS) whose definition, namely
\be \lab{cksdef}
\left(\g_\m D_\n +\g_\n D_\m - \frac12 g_{\m\n} \slashed{D}\right)\e =0,
\ee
is compatible with the Majorana property.  Contracting with $\g^\m$, one finds the equivalent condition
\be \lab{cks2}
\left(D_\m - \frac14 \g_\m\slashed{D}\right)\e =0\,.
\ee
One can show that the real and imaginary parts of any complex spinor in dS$_4$ that satisfies \reef{dsksp} are Majorana conformal Killing spinors (see appendix \ref{CKSAp}).  We will obtain the full set of Majorana spinors in a simpler way, but this requires a choice of metric on dS$_4$.

Many coordinate systems are useful to discuss physics in de Sitter space. Since conformal symmetry is central for us we use the conformally flat  ``Poincar\'e patch metric" (obtained from planar  coordinates by the transformation $x^0 \equiv- a \exp(-t/a)$):
\be  \lab{dsmetric}
ds^2 = - dt^2+ e^{2t/a} d\vec{\mathbf{x}}^2  =  \frac{a^2}{(x^0)^2} \left[ -(dx^0)^2 + d\vec{\mathbf{x}}^2\right]\,.
\ee
Its structure is similar to  the Poincar\'e patch of anti-de Sitter space. The time coordinate $x^0\in[-\infty,0]$ is chosen so that spatial volumes increase with increasing $x^0$.

In flat spacetime the most general CKS, given, for example,  by Wess and Zumino \cite{Wess:1974tw}, is  $\h_0 +  x^{\hat{\m}}\g_{\hat{\m}}\, \h_1$,   where $\h_0,~\h_1$ are constant Majorana spinors.   The simplest way to obtain the CKS's for dS$_4$ is to use the Weyl transformation from flat spacetime to define
\be  \lab{cksds}
 \epsilon (x)  = \frac{1}{\sqrt{-x^0/a}} \left(\h_0 +x^\m\g_{\hat{\m}}\,  \h_1\right)\,.
\ee
In the frame  
$e^{\hat{\m}}= -\frac{a}{x^0} dx^\m$, 
the spinor covariant derivatives are:
\be \lab{covder}
D_0 \e \equiv \pa_0 \e   \qquad\quad  D_i \e= \left(\pa_i + \frac{1}{2 x^0}\g_{\hat{i}}\g_{\hat{0}}\right)\e\,~.
\ee
It is straightforward to check that spinors \reef{cksds} satisfy \reef{cks2}.  Thus we have a basis containing 8 real supercharges.  



One can calculate the weights $\tfrac14 D_\r (\bar\e'\g^\r\e) $ of all bilinears in the basis \reef{cksds}  and learn that they do not vanish.
Thus, when $\e',~\e$ are CKS's, the SUSY commutator  $[\d_{\e'}\,,\,\d_\e]$ contains the CKV's of dS$_4$, albeit with admixtures of KV's.  As we will argue below, this leads to a supercharge anti-commutator of the schematic form 
\be \lab{qqdag}
\sum_\a\{Q_{\a}, Q^{\dagger\a}\} = - \int d^3x \sqrt{-g} K_\n(x) T^{0\n}\,,
 \ee
 in which $K^\n$ is a future directed time-like CKV.  Such conformal Killing charges are conserved formally, i.e. if  boundary conditions permit, and if the stress tensor is conserved and traceless.  
 Positivity holds if the stress tensor of the theory satisfies the dominant energy condition.  We elaborate on this point in section \ref{Positivitysec}.
  
The conformal group of dS$_4$ is $SO(4,2)$.  Its Lie algebra is realized by the ten KV's of the isometry group $SO(4,1)$ plus the five CKV's of the coset $SO(4,2)/SO(4,1)$.   The CKV's are translations of the embedding coordinates $Y^A$ of the hyperboloid.  It is worthwhile to display them as vector fields on the surface. In patch coordinates the five CKV's are:
\begin{enumerate}[i.] 
\item Translation of patch time  $K_0=\pa_0$.  This is clearly  future-directed and time-like.
\item The Lorentz boost   $K_i = \d_{ij} x^j \pa_0 + x^0 \pa_i$ which is not future-directed.
\item The special conformal $K_2 = ( \vec{\mathbf{x}}\cdot \vec{\mathbf{x}} +(x^0)^2) \pa_0 +2x^0  \vec{\mathbf{x}}\cdot \pa_{\vec{\mathbf{x}}}$,   also future-directed and time-like.
\end{enumerate}
In dS$_4$ they are genuine CKV's with non-vanishing weights $\tfrac14D_\m K^\m$.  Readers are invited to compute these weights.  The CKV's  $K_0$ and $K_2$ are the main contribution to \reef{qqdag}.

The Lie brackets $[CKV, CKV']$ are linear combinations of the 10 KV's,  namely
3 space transformations,  3 rotations, 3 spatial special conformal transformations  and the scale transformation $K_s = x^\m \pa_\m$.  


\section{The basic $\cn =1$ SCFT's in dS$_4$}

\subsection{Non-abelian $\cn=1$ SYM theory}

The fields are $A_\m^a,\,\l^a,\,D^a$, where $a$ is the index of the gauge group.
The action is\footnote{We use the conventions of Ch. 6 of \cite{Freedman:2012zz}, except that SUSY parameters are multiplied by $\sqrt2$.}
\be \lab{ssym}
S_{gauge}=-\int d^4x \sqrt{-g} \bigg[\frac14 F^a_{\m\n}F^{a\m\n}+\frac12\bar\l^a\slashed{D}\l^a +\frac12 D^aD^a\bigg]\,.
\ee
This is just the flat space action extended by  minimal coupling to the dS$_4$ metric and spin connection. The same holds for the 
SUSY variations:   
 \bea  \lab{gaugemulttrfs}
 \d A^a_\m &=& - \frac{1}{\sqrt2}\bar\e \g^\m\l^a\\
\d\l^a &=& \frac{1}{\sqrt2}\left[\frac12 \g^{\r\s}F^a_{\r\s} + i \g_5 D^a\right]\e\\
\d D^a &=& \frac{1}{\sqrt2}\bar\e \g_5 \slashed{D}\l^a\,.
\eea

It is now very easy to show that the action  is invariant if $\e$ is a CKS and thus satisfies \reef{cksdef}.  We simply ``covariantize" the flat space calculation of \cite{Freedman:2012zz} and show for a general $\e(x)$ that
\be
\d S \,=\, -\frac{1}{2\sqrt2} \int d^4x\sqrt{-g}\bar\e\overleftarrow{D}_\m \g^{\r\s}\g^\m F^a_{\r\s}\l^a\,.
\ee 
Note that $\bar\e\overleftarrow{D}_\m$ is contracted with the conserved supercurrent.
Now choose $\bar{\e}(x)$ to  be a CKS and use \reef{cksdef} to write
\be
\d S \,=\, -\frac{1}{8\sqrt{2}}\int d^4x\sqrt{-g}\bar\e\overleftarrow{\slashed{D}}( \g_\m\g^{\r\s}\g^\m) F^a_{\r\s}\l^a\,.
\ee
But $\g_\m\g^{\r\s}\g^\m= (d-4)\g^{\r\s}\to 0$ for $d=4$.  
The first $\cn=1$ de Sitter SUSY theory is thus established!

\subsection{The superconformal chiral multiplet}

The $\cn=1$ chiral multiplet contains the  fields  $z, \,\chi,\,F$. The extension of the flat space kinetic action to dS$_4$ includes the conformal coupling $-R\bar{z}z/6 = -2  \bar{z}z/a^2$ and is given by
\be  \lab{skinchrl}
S_{\rm kin} \,=\,\int d^4x\sqrt{-g}\bigg[-g^{\m\n}\pa_\m \bar z \pa_\n z- \frac12\bar\chi \slashed{D}\chi -\frac{2}{a^2}\bar{z}z+ \bar F F \bigg]
\ee
It is invariant under the transformation rules:
\bea \lab{chimulttrfs}
\d z &=& \bar\e P_L \chi~, ~~~~~~~~\d \bar{z} = \bar\e P_R \chi~,\\
\d \chi &=& P_L\left[\slashed{D}z +F  + \frac12 z \slashed{D}\right]\e+P_R\left[\slashed{D}\bar{z} +\bar{F}  + \frac12 \bar{z} \slashed{D}\right]\e~,\label{improved}\\
\d F &=& \bar\e \slashed{D} P_L\chi~,~~~~~ \d {\bar F} = {\bar \eps}{\slashed D} P_R  \chi~.
\eea
The coefficient of the term proportional to $ \slashed{D}\e $ in (\ref{improved}) is fixed by the requirement that the sum of all terms  transforms with weight 3/2 under the Weyl transformation  $g_{\m\n} \to g'_{\m\n} = e^{2\s(x)}g_{\m\n}. $
Under this transformation  $\e  \to \e' = e^{\s/2}$,  $z\to z' = e^{-\s}z,~~F\to e^{-2\s}F$,  $ \chi \to \chi' = e^{-3\s/2}\chi.$   One may check to see that $\pa_\m\s$ cancels in the sum of the three terms in $\d P_L\chi$. 

The transformations  \reef{chimulttrfs} are generated by the improved supercurrent
\be 
{\cal J}^\m  =P_L\left[( \slashed{D}\bar z -F)\g^\m\chi  +\frac23\g^{\m\n}D_\n(\bar z\chi)\right] +P_R\left[( \slashed{D}z -\bar F)\g^\m\chi  +\frac23\g^{\m\n}D_\n( z\chi) \right]\,,
\ee  
which is conserved and $\g$-traceless.

\subsection{Superpotentials}

The chiral multiplet considered above is a theory of free fields.  In flat space the Wess-Zumino model with cubic superpotential $W= z^3/3$ is superconformally invariant, so it should have an extension to  de Sitter space. We begin the discussion more  broadly and consider a general holomorphic superpotential $W(z)$.  The de Sitter covariant extension of its flat space interaction term is  
\be \lab{sw}
S_W = \int d^4x \sqrt{-g}[ F W' - \frac12 (\bar\chi P_L \chi) W'']\,.
 \ee
It is easy to verify that its variation under the transformations  \reef{chimulttrfs} is
\be \lab{delsw}
\d S_W =\int d^4x \sqrt{-g}\bar\chi(W'- z W''/2)P_L\, \slashed{D}\e\,.
\ee
This vanishes only for a cubic superpotential.
The complete interacting theory is then specified by the sum $S= S_{\rm kin} + S_W +S_{\bar W}$ with $W$ and $\bar{W}$ cubic in $z$ and $\bar{z}$ respectively.  

Let us look briefly at the possibility of generating supersymmetry using Killing spinors $\e(x)$ that 
satisfy \reef{dsksp} and thus also $\slashed{D}\e= \frac{2i}{a} \e$.  The kinetic action \reef{skinchrl} is invariant because it is manifestly real and the real and imaginary parts of $\e(x)$ are each conformal Killing spinors. Although the variation $\d S_W$ does not vanish for general $W(z)$,  the sum $S_W + \D S_W$ is invariant if $\D S_W$ is chosen as the scalar term
\be 
\D S_W = -\frac{3i}{a} \int d^4x \sqrt{-g}(W - zW'/3)\,.
\ee
Notice that $\D S_W$ vanishes when $W$ is cubic. However, as an example of the problems with this construction for general $W$, consider the mass term $W= m z^2/2$ and its conjugate $\bar W =m \bar z^2/2$.  After elimination of the auxiliary fields one finds a non-hermitean scalar potential (note $z = (A+iB)/\sqrt2$)
\be
V = \frac12 (2/a^2 +m^2 + im/a)A^2 + \frac12 (2/a^2 +m^2 - im/a)B^2
\ee
with complex masses. 
A further pathology is that the SUSY variations $\d z = \bar{\e}P_L\chi$ and $\d\bar z = \bar\e P_R \chi$ are not related by complex conjugation if $\e$ is a complex Killing spinor.   

\subsection{Chiral multiplet with gauge interactions}

Generalizing the $\cn=1$ chiral multiplet to include gauge interactions is a  straightforward exercise that follows from the flat space case. 
The fields $(z, \,\chi,\,F)$ are now taken to transform in a representation of the gauge group with generators $t_a$.
The kinetic term is exactly as in (\ref{skinchrl}), except that the derivatives are replaced by the usual gauge covariant derivatives
\be
\pa_\mu z \rightarrow \pa_\mu z + t_a A_\mu^a z,~~~~D_\mu P_L \chi \rightarrow D_\mu P_L \chi + t_a A_\mu^a P_L \chi,~~~~ 
D_\mu P_R \chi \rightarrow D_\mu P_R \chi + t^*_a A_\mu^a P_R \chi ~.
\ee 
The transformation rules
 \bea \lab{chimulttrfs}
\d z &=& \bar\e P_L \chi\\
\d P_L\chi &=& P_L\left[\slashed{D}z +F  + \frac12 z \slashed{D}\right]\e\\
\d F &=& \bar\e \slashed{D} P_L\chi -\sqrt{2} \bar \e P_R \lambda^a t_a { z}
\eea
include the usual gauge covariant derivatives as well as an additional term in $\d F$ involving $\lambda$ and $z$.
In addition, the following coupling term 
\be
S_{coupling} 
=
\int d^4 x\sqrt{-g} \left[
-\sqrt{2} \left(\bar\lambda^a \bar z t_a P_L \chi - \bar \chi P_R t_a z \lambda^a\right) + i D^a \bar z t_a z\right]
\ee
must be added to the action to cancel the SUSY variation of the gauge fields appearing in the kinetic action.
The proof of invariance of the full action
\be
S= S_{gauge} + S_{kin} + S_{coupling}
\ee
then proceeds much as in the flat space case.

\section{The SUSY algebra for conformal Killing spinors}

In this section we study the SUSY algebra of the $\cn=1$ chiral and gauge multiplets in dS$_4$.  We assume that $\e(x),~\e'(x)$ are conformal Killing spinors
that satisfy \reef{cks2}.  We define  $K^\m = \bar{\e}\g^\m\e'$ and $\k = \frac14(\bar \e\g_5\slashed{D}\e' - \bar \e'\g_5\slashed{D}\e)$ .  For the chiral multiplet,   the commutator of the variations  of  \reef{chimulttrfs} gives

\begin{align}
&[\d',\d]z = K^\m D_\m z+(\tfrac{1}{4}D_\m K^\m)z+\k z\label{scalaralg}\\
&[\d',\d]P_L\chi =K^\m D_\m\,P_L\chi+\frac{3}{2}(\tfrac{1}{4}D_\m K^\m)P_L\chi+\frac{1}{8}\left(D_\m K_\n-D_\n K_\m\right)\g^{\m\n}P_L\chi-\frac{1}{2}\k\g_*P_L\chi\label{fermalg}\\
&[\d',\d]F=K^\m D_\m F+2(\tfrac{1}{4} D_\m K^\m)F-2\k F
\end{align}

For the  gauge multiplet the analogous computation from  \reef{gaugemulttrfs} leads  to
\begin{align}
&[\d',\d]A_\r=K^\m F_{\m\r}\\
&[\d',\d]\lambda =K^\m D_\m \lambda+\frac{3}{2}(\tfrac{1}{4} D_\m K^\m)\lambda+\frac{1}{8}\left( D_\m  K_\n- D_\n K_\m\right)\g^{\m\n}\lambda+\frac{3}{2}\k\g_*\lambda\\
&[\d',\d]D=K^\m D_\m D+2(\tfrac{1}{4} D_\m K^\m)D
\end{align}

The structure and the coefficients agree with the commutators of flat space superconformal algebra  compiled in \cite{Wess:1974tw}.  The first term is the Lie derivative that effects the diffeomorphism generated by the conformal Killing vector $K^\m$.  It is accompanied by the conformal weight term whose coefficients are the scale dimensions 1, 3/2, 2 respectively for the fields $z, \chi,~F$.  There is  a local Lorentz transformation on $\chi$.  The last term is the $U(1)_R$ transformation of the superconformal algebra.  (The coefficients of $\k$ satisfy the expected relations between the R-charges of the various fields, but they are scaled by the factor 3/2 from the conventional values.)  Similar comments apply to the gauge multiplet. 

\section{The supercharge algebra and unitarity}\label{Positivitysec}

In this section we express the SUSY algebra in terms of conserved charges and derive \reef{poscom} at the classical level.  The Noether charge associated with a covariantly conserved vector current $J^\mu(x)$ is given by
\be\lab{charge}
Q= - \int d^3x\sqrt{\gamma}~n_\m J^\m  =  \int d^3x\sqrt{-g}J^0 ~.
\ee
In the first expression $d^3x\sqrt{\g}$ is the volume element of the spatial 3-volume orthogonal to the future-pointing timelike unit normal $n^\m$.  For the Poincar\'e patch metric \reef{dsmetric} of dS$_4$,  this volume may be taken to be the spatial slice of constant conformal time $x^0$ (which is negative in our conventions).   The
normal vector is $n^\m =  (n^0=-x^0/a,0,0,0)$.   Then $n_\m=n_0=a/x^0$ and $\sqrt{\g}n_0=-\sqrt{-g}$ where $g$ is the determinant of the full metric of dS$_4$.
Thus we obtain the second expression for the charge.  This charge is independent of the time $x^0$ provided that the current falls off sufficiently fast at spatial infinity.

For a conformal Killing vector $K_\n$,   the current $J^\m = - K_\n \Theta^{\m\n}$ is covariantly conserved provided that the stress tensor $\Theta^{\m\n}$ is conserved and traceless. 
For a conformal Killing spinor $\e(x)$,  the current  $J^\m = \bar\e^\a {\cal J}^\m_\a$ 
is conserved if the supercurrent ${\cal J}^\m$ is covariantly conserved and $\g$-traceless,  i.e.  $\g_\m {\cal J}^\m =0.$  Thus we deal with the conformal Killing charges and supercharges
\bea\lab{scharges}
Q[K]  &=& - \int d^3x\sqrt{-g} K_\n \Theta^{0\n}\\
\bar\e^\a Q_\a &=&  \int d^3x\sqrt{-g} \bar\e^\a {\cal J}^0_\a = \bar Q^\a \e_a~.
\eea  
Then the operator algebra statement that corresponds to the commutators of variations in the previous section is
\be\lab{alg1}
[\bar\e^{'\a} Q_\a,   \bar Q^\b \e_\b]  =  -i \,Q[ \bar\e' \g_\n \e]+\ldots \,.
\ee

We now need to be more specific and strip off the constant Grassmann parameters
$\bar\h,~ \h$ in \reef{alg1}.  We refer to the specific spinors of \reef{cksds} and define the two spinor supercharges:
\bea
\bar\h^\a_1 Q_{1\a}  &\equiv&   \bar\eta^\a_1\int d^3x\sqrt{-g}\, (\bar{S}_1(x))_\a{}^\b {\cal J}^0_\b\\
\bar\h^\a_2 Q_{2\a}     &\equiv&  \bar\h^\a_2\int d^3x\sqrt{-g}\, (\bar{S}_2(x))_\a{}^\b {\cal J}^0_\b\\
S_1(x) &\equiv& \frac{1}{\sqrt{-x^0/a}}  \qquad S_2(x) \equiv \frac{1}{\sqrt{-x^0/a}}(\g_{\hat{\r}} x^\r)\,,  \lab{s1s2}
\eea
where  $S_1(x),~S_2(x)$  are $4\times 4$ matrices whose Dirac adjoints $\bar S_i = i\g^{\hat{0}}(S_i)^\dagger i\g^{\hat{0}}$ appear in the supercharges.   Then \reef{alg1} can be replaced by the supercharge anti-commutator
\be
\{Q_{i\a}\,,\,\bar Q^{\b}_j\}   = -i \int d^3x\sqrt{-g} (\bar{S}_i(x)\g_\n S_j(x))_\a{}^\b \Theta^{0\n}\,.
\ee

Since we are interested in deriving \reef{poscom},   we multiply from the right by $i\g^{\hat{0}}$ and trace,  and we choose  $i=j$ with no sum.  We find the traced anti-commutator
\be  \lab{alg2}
\sum_\alpha \{Q_{i\a}\,,\, Q^{\dagger\a}_i\}  = \int d^3x\sqrt{-g} {\rm Tr}(\bar{S}_i(x)\g_\n S_i(x)\g^{\hat{0}})\Theta^{0\n}\,.
\ee
For $i=1,2$, the traces are
\bea   \lab{posckvs}
  {\rm Tr}(\bar{S}_1(x)\g^\n S_1(x)\g^{\hat{0}})  &=&  -4 (1,0,0,0)  = -4 K_0^\n \\
 {\rm Tr}(\bar{S}_2(x)\g^\n S_2(x)\g^{\hat{0}})  &=& -4( 2x^0 x^\n - x\cdot x \eta^{\n 0})= -4 K_2^\n\,.  \lab{poscks2}
 \eea
 In each case we find one of the future pointing conformal Killing vectors described at the end of Sec. 2.   So the general form of \reef{alg2}  becomes
\be \lab{alg3}
\tfrac14\sum_\a\{Q_{i\a}\,,\, Q^{\dagger\a}_i\}  =- \int d^3x\sqrt{-g}  K_\n\Theta^{0\n}\,.
\ee
in which $K^\n$ is the conformal Killing vector for time translation or time-like special conformal transformation in the Poincar\'e patch.  

We now argue that the the operator on the right hand side of \reef{alg3} is positive in the classical approximation in which we include  only the bosonic contribution to the stress tensor of  our basic $\cn=1$ theories.  Given \reef{charge}, it is the integral of $n_\m K_\n \Theta^{\mu\nu}$ that must be non-negative.  Let's first look into the stronger condition of positivity of the integrand.   We note that the standard dominant energy condition requires that
\be  \lab{domenergy}
A_\m B_\n  \Theta^{\mu\nu}  \ge 0
\ee
where $A^\m,~B^\n$ is any pair of future pointing timelike or null vectors.  The dominant energy condition is well known \cite{Rendall} for the electromagnetic   field and the proof is immediately applicable to Yang-Mills.  
Dominant energy is also valid for the  canonical  stress tensor of the free scalar field \cite{Rendall},  but it does not hold for conformally coupled scalar of the chiral multiplet  \cite{Flanagan:1996gw,Visser:1999de,Barcelo:2000zf}.  

Let us look at the integral in \reef{alg3} for a real conformally coupled scalar field $\phi$ and try to show it is positive.  The improved stress tensor, which is both conserved and traceless, is:
\begin{align}
\Theta_{\m\n}&= D_\m\phi D_\n\phi-\frac{1}{2}g_{\m\n}( D_\r\phi D^\r\phi)-\frac{1}{6}\left( D_\m D_\n-g_{\m\n}\Box-R_{\m\n}+\frac{1}{2}g_{\m\n}R\right)\phi^2\\
&=\frac{2}{3} D_\m\phi D_\n\phi-\frac{1}{6}g_{\m\n}( D_\r\phi D^\r\phi)-\frac{1}{3}\phi D_\m D_\n\phi+\frac{1}{6a^2}g_{\m\n}\phi^2
\end{align}
where in the second line we used $\Box\phi=(2/a^2)\phi$, $R_{\m\n}=\tfrac{1}{4}g_{\m\n}R$ and $R=12/a^2$.  We work in the Poincar\'e patch and consider the conformal Killing charge
\begin{align} \lab{ckc}
Q[K]=&-\int d^3x\sqrt{-g}~ \Theta^0_{~\n}K^\n~\\
=&-\int d^3x\sqrt{-g}\bigg[K^0\left(\frac{2}{3} D^0\phi D_0\phi-\frac{1}{6}( D_\r\phi D^\r\phi)-\frac{1}{3}\phi D^0 D_0\phi+\frac{1}{6a^2}\phi^2\right)\nonumber\\
&~~~~~~~~~~~~~~~~~~~~~~~~~~~~~~~~~~~~~~~~~~~~~~~~~~~~~~~~+K^i\left(\frac{2}{3} D^0\phi D_i\phi-\frac{1}{3}\phi D_i D^0\phi\right)\bigg]\,.
\end{align}
The first step is to use the equation of motion in the form 
\be
 D^0 D_0\phi=\left(\frac{2}{a^2}- D^i D_i\right)\phi~
\ee
to rewrite the integrand in  \reef{ckc} as
\begin{multline}
Q[K]=-\int d^3x\sqrt{-g}\bigg[K^0\left(\frac{1}{2} D^0\phi D_0\phi-\frac{1}{6} D^i\phi D_i\phi+\frac{1}{3}\phi D^i D_i\phi-\frac{1}{2a^2}\phi^2\right)\\
+K^i\left(\frac{2}{3} D^0\phi D_i\phi-\frac{1}{3}\phi D_i D^0\phi\right)\bigg]~.
\end{multline}
We assume that the fields on the time slice decay fast enough at spatial infinity to allow partial integration without picking up boundary contributions.  We integrate  the 3rd term in the first line and the 2nd term in the second line above to obtain
\begin{multline}
Q[K]=-\int d^3x\sqrt{-g}\bigg[K^0\left(\frac{1}{2} D^0\phi D_0\phi-\frac{1}{2} D^i\phi D_i\phi-\frac{1}{2a^2}\phi^2\right)+K^i\left( D^0\phi D_i\phi\right)\\
-\frac{1}{3} D_i K^0\left(\phi D^i\phi\right)+\frac{1}{3} D_iK^i\left(\phi D^0\phi\right)\bigg]~.
\end{multline}
Next we substitute  $ D^0\phi=-\tfrac{(x^0)^2}{a^2} D_0\phi$ and $ D^i\phi=\tfrac{(x^0)^2}{a^2} D_i\phi$, obtaining
\begin{multline}\label{massagedQK}
Q[K]=\int d^3x\sqrt{-g}\frac{(x^0)^2}{a^2}\bigg[\frac{K^0}{2}\left( D_0\phi D_0\phi+ D_i\phi D_i\phi+\frac{1}{(x^0)^2}\phi^2\right)+K^i\left( D_0\phi D_i\phi\right)\\
+\frac{1}{3} D_i K^0\left(\phi D_i\phi\right)+\frac{1}{3} D_iK^i\left(\phi D_0\phi\right)\bigg]~.
\end{multline}

We now apply \reef{massagedQK} to the two positive timelike CKV's in (\ref{posckvs}-\ref{poscks2}).  For the generator of  translations of conformal time, $K_0=\pa_0$. The CKV equation implies  $ D_i K_0^0=0$ and $\frac{1}{3} D_i K_0^i=-1/x^0$. We regroup terms in \reef{massagedQK} and find 
\begin{align}\label{conftimepos}
Q[K_0]&=\int d^3x\sqrt{-g}\bigg[\frac{1}{2a^2}\left((x^0)^2(D_{\vec{\mathbf{x}}}\phi)^2+\left(\phi-x^0 D_0\phi\right)^2\right)\bigg]~
\end{align}
which is non-negative.  

The special conformal CKV, $K_2 = ( \vec{\mathbf{x}}\cdot \vec{\mathbf{x}} +(x^0)^2) \pa_0 +2x^0\,  \vec{\mathbf{x}}\cdot \pa_{\vec{\mathbf{x}}}$~, satisfies
$ D_i K_2^0 = 0$ and $\frac{1}{3} D_iK_2^i = -x\cdot x/(x^0)$~, where $x\cdot x=-(x^0)^2+ \vec{\mathbf{x}}\cdot \vec{\mathbf{x}}$. In this case one can verify that 
\begin{multline}\label{specialconfpos}  
Q[K_2]=\int d^3x\sqrt{-g}\bigg[\frac{ \vec{\mathbf{x}}\cdot \vec{\mathbf{x}}+(x^0)^2}{2a^2}\bigg\lbrace(x^0)^2\left(D_{\vec{\mathbf{x}}}\phi+\frac{2  x^0\,\vec{\mathbf{x}}}{ \vec{\mathbf{x}}\cdot \vec{\mathbf{x}}+(x^0)^2} D_0\phi\right)^2\\
+\left(\phi-\frac{ x^0 x\cdot x}{ \vec{\mathbf{x}}\cdot \vec{\mathbf{x}}+(x^0)^2} D_0\phi\right)^2\bigg\rbrace\bigg]~,
\end{multline}
again non-negative.

The results  (\ref{conftimepos}) and (\ref{specialconfpos}) actually have the common form 
\be\label{gencharge}
Q[K]=\int d^3x\sqrt{-g}\bigg[\frac{K^0}{2a^2}\left((x^0)^2\left( D_i\phi+\frac{K^i}{K^0} D_0\phi\right)^2+\left(\phi+\frac{1}{3} D_iK^i\frac{(x^0)^2}{K^0} D_0\phi\right)^2\right)\bigg]~.
\ee
This form of the integrated conformal Killing charge matches with (\ref{massagedQK}) for any CKV that satisfies
\be
 D_i K^0=0\quad\text{and}\quad (K^i)^2+(x^0)^2\left(\frac{1}{3} D_i K^i\right)^2=(K^0)^2~.
\ee
These conditions are satisfied by the two CKV's that appear in the traced de Sitter SUSY algebra \reef{alg3}.   They are also satisfied by the three Lorentz boost CKV's  discussed in Sec. 2,  namely  $K_{(i)}^0 = \d_{ij} x^j, ~  K_{(i)}^j = x^0 \d^j_i$,  although their integrated charges are not necessarily positive.

So far we have considered the SUSY anti-commutator $\{Q_\a, Q^{\dagger\a}\}$ summed over spinor components.   We would like to show that each individual diagonal term of  $\{Q_\a, Q^{\dagger\b}\}$  is classically positive.  We  summarize our progress on this question. In the usual Weyl representation of the Dirac algebra (and in the Majorana representation of Sec 3.3.1 of  \cite{Freedman:2012zz}),  the diagonal terms take a  simple form.  The CKV $K_0^\n$ in \reef{posckvs} is replaced by $(1,0,0,\pm 1)$, the sum of the conformal time translation and a space translation.   Similarly,
$K_2^\n$ is replaced by $v_\m (2x^\m x^\n -x\cdot x \h^{\m\n})$ which describes a special conformal transformation in the null direction $v_\m = (1,0,0,\pm 1)$.  In both cases these are future-pointing null vectors,
so the dominant energy condition implies positivity for the gauge multiplet.   For the chiral multiplet we were not able to extend the analytic proof of positivity above.

\section{Propagator Ward Identities and the Vacuum}

In a supersymmetric field theory there are Ward identities that relate the propagators of boson and fermion fields.  We will argue in this section that superconformal Ward identities select the Bunch-Davies vacuum state which is the unique conformal invariant state among the well known one-parameter family of $SO(4,1)$ invariant vacua.

We begin with a brief discussion of the propagators\footnote{We do not specify the time ordering prescription.  See   \cite{Spradlin:2001pw} for more information.}  of a massive scalar field which satisfy   
\be\label{scalarprop}
(\Box-m^2)G(x,y)=\frac{ \delta(x,y)}{\sqrt{-g}}\,.
\ee
A de Sitter invariant solution of this equation
 will be a function of $P(x,y)$ \cite{Spradlin:2001pw} defined in terms of the embedding coordinates of the hyperboloid by:
\begin{equation}
P(x,y)=\eta_{AB}Y^A(x)Y^B(y)~.
\end{equation}
The geodesic distance between two points $D(x,y)$ in dS$_4$ is given by $D(x,y)=\arccos P(x,y)/a^2$. We use instead the chordal distance variable $u\equiv 1-P/a^2$.  In Poincar\'e patch coordinates, $u$ takes the simple form
\be
u=\frac{\h_{\m\n}(x-y)^\m(x-y)^\n}{2x^0 y^0}~.
\ee
One can convert \reef{scalarprop} into the hypergeometric equation
\be  \lab{hypergeo}
u(2-u) G''(u) + 4(1-u) G'(u) - (ma)^2 G(u) = 0\,.
\ee
The general solution to (\ref{hypergeo}) is given by:
\begin{equation}\label{scalar2point}
G(u)=\frac{\Gamma(h_+)\Gamma(h_-)}{16\pi^2a^2}\left[c_1\; _{2}F_1\left(h_+,h_-,2,1-\tfrac{u}{2}\right) + c_2\;_{2}F_1\left(h_+,h_-,2,\tfrac{u}{2}\right) \right]~,
\end{equation}
where
\begin{equation}
h_\pm=\frac{3}{2}\pm\sqrt{\frac{9}{4}-m^2 a^2}~.
\end{equation}
The first term in (\ref{scalar2point}) has the expected singularity for null separated points $Y^A(x)$ and $Y^B(y)$,  i.e. $(Y^A(x)-Y^A(y))^2 =0$. The second term is singular when one point is null separated from the antipodal reflection of the other, i.e. when $(Y^A(x)+Y^A(y))^2 =0$.
In de Sitter space this unphysical term cannot be discarded because two antipodal points are separated by a horizon, and the singularity is  undetectable to geodesic observers.  Henceforth we take $c_1=1$ in order to normalize to the $\d$-function in \reef{scalarprop}, and simply replace $c_2 \to c.$  The one-parameter family of de Sitter invariant propagators in \reef{scalar2point} then corresponds to the family of  Mottola-Allen or $\alpha$-vacua of a field theory on dS$_4$ (see e.g. \cite{Spradlin:2001pw,Anninos:2012qw,Mottola:1984ar,Allen:1985ux,Bousso:2001mw}).  Of these the choice $c=0$ gives the Bunch-Davies vacuum. 

%

The hypergeometric solution   (\ref{scalar2point}) simplifies significantly at the conformally coupled mass point $m^2=2/a^2$ where we find: 
\be\label{scalar2pointconf}
G(u)=\frac{1}{8\pi^2a^2}\left(\frac{1}{u}+\frac{c}{2-u}\right)~.
\ee
In this case the Bunch-Davies choice is the unique conformally invariant vacuum. One simple way to see this is to notice that, when $c=0$, $G(u)$ is related by to the flat space propagator by the factor $(x^0 y^0)/a^2$,  determined by the Weyl transformation
$\phi'(x) = -(x^0/a) \phi(x)$  of the field to the Poincar\'e patch  (see \reef{dsmetric}).  Alternatively, this can be seen by noting that the condition that $x$ and $y$ are antipodally separated points is not preserved by a conformal transformation on de Sitter space (though, of course the condition that $x$ and $y$ are coincident is preserved). 

\subsection{Ward Identities}
We are now ready to discuss propagator Ward identities for the chiral multiplet. Using equations~\reef{chimulttrfs} and noting that $ \d\langle\chi(x)z(y)\rangle= \d\langle\chi(x)\bar{z}(y)\rangle=0$ it is easy to derive a consistency condition satisfied by the propagator $\langle\chi(x)\bar{\chi}(y)\rangle$. In fact we use this to \emph{define} $\langle\chi(x)\bar{\chi}(y)\rangle$.  We consider the SUSY 
transformations \reef{chimulttrfs} for the 
general CKS $\epsilon(x)$ of \reef{cksds}.  We use \reef{s1s2} and note that
$\slashed{ D}S_1=-\tfrac{2}{a}S_{1} \gamma_{\hat{0}}$ and $\slashed{ D}S_2=+\tfrac{2}{a}S_{2} \gamma_{\hat{0}}$. This allows us to derive two independent expressions for $\langle\chi(x)\bar{\chi}(y)\rangle$, namely: 
\begin{align}\label{consistcond1}
 \langle\chi(x)\bar{\chi}(y)\rangle &=-\slashed{ D}_x\langle\bar{z}(x)z(y)\rangle S_1(x)S_1^{-1}(y)+\frac{1}{a}\langle\bar{z}(x)z(y)\rangle S_1(x)\gamma_{\hat{0}}S_1^{-1}(y)\\
&=-\slashed{ D}_x\langle\bar{z}(x)z(y)\rangle S_2(x)S_2^{-1}(y)-\frac{1}{a}\langle\bar{z}(x)z(y)\rangle S_2(x)\gamma_{\hat{0}}S_2^{-1}(y)\label{consistcond2}
\end{align}
where $\langle\bar{z}(x)z(y)\rangle$ is the scalar two-point function given in (\ref{scalar2pointconf}). A calculation (Mathematica!) then shows that the two  conditions (\ref{consistcond1}-\ref{consistcond2}) are mutually consistent  only for the Bunch-Davies vacuum where $c=0$. This is natural since only this vacuum is conformal invariant, but it is satisfying to see how conformal SUSY forces this choice. 

Using (\ref{consistcond1}-\ref{consistcond2}), it is easy to show that
\be
 \langle\chi(x)\bar{\chi}(y)\rangle=\frac{1}{8\pi^2 a^3\, u^2}\frac{(x-y)^\m}{\sqrt{x^0 y^0}} \gamma_{\hat{\m}}~.
\ee
Notice again that this propagator is related to the flat space propagator by the factor of $((x^0 y^0)/a^2)^{3/2}$,  corresponding to the Weyl transformation $\psi'(x) = (-x^0/a)^{3/2} \psi(x).$

It would be interesting to consider the propagator Ward identities for the gauge multiplet.  However,  the analysis becomes more complicated because the consistency conditions include a new term due to the SUSY variation of the gauge-fixing term in the action.

\section{Hawking Temperature and dS SUSY}

We now comment on the relationship between Hawking radiation and de Sitter supersymmetry.
An observer in de Sitter space will observe a bath of thermal particles emitted from the de Sitter horizon at the Hawking temperature
\be\
T_{H} = {1\over 2\pi a}~.
\ee
One might worry that the non-zero Hawking temperature of de Sitter space breaks supersymmetry, because in flat space 
a finite temperature breaks supersymmetry.
We will now argue that this is not the case.

We begin by recalling the derivation of Hawking radiation in de Sitter space.  The Hilbert space of quantum field theory in a fixed de Sitter background is 
built on the Bunch-Davies 
vacuum state $|0\rangle$.  As we saw in Section 6, the Bunch-Davies state is consistent with the SUSY Ward identities.  That is to say, the state $|0\rangle$ has unbroken supersymmetry.
We consider an inertial observer in de Sitter space who makes observations in this vacuum state. 
To this observer, we can associate a static patch of de Sitter space, with coordinates
\be\label{causal}
{ds^2 \over a^2} = -({1-r^2})dt^2 + {dr^2 \over 1-r^2} + r^2 d\Omega^2  
\ee
The observer sits at $r=0$ and the de Sitter horizon is at $r=1$.  The coordinate patch $0\le r<1$ covers that part of de Sitter space which is causally accessible to the observer. 
\begin{figure}
\centering{
\includegraphics[height=6.5cm]{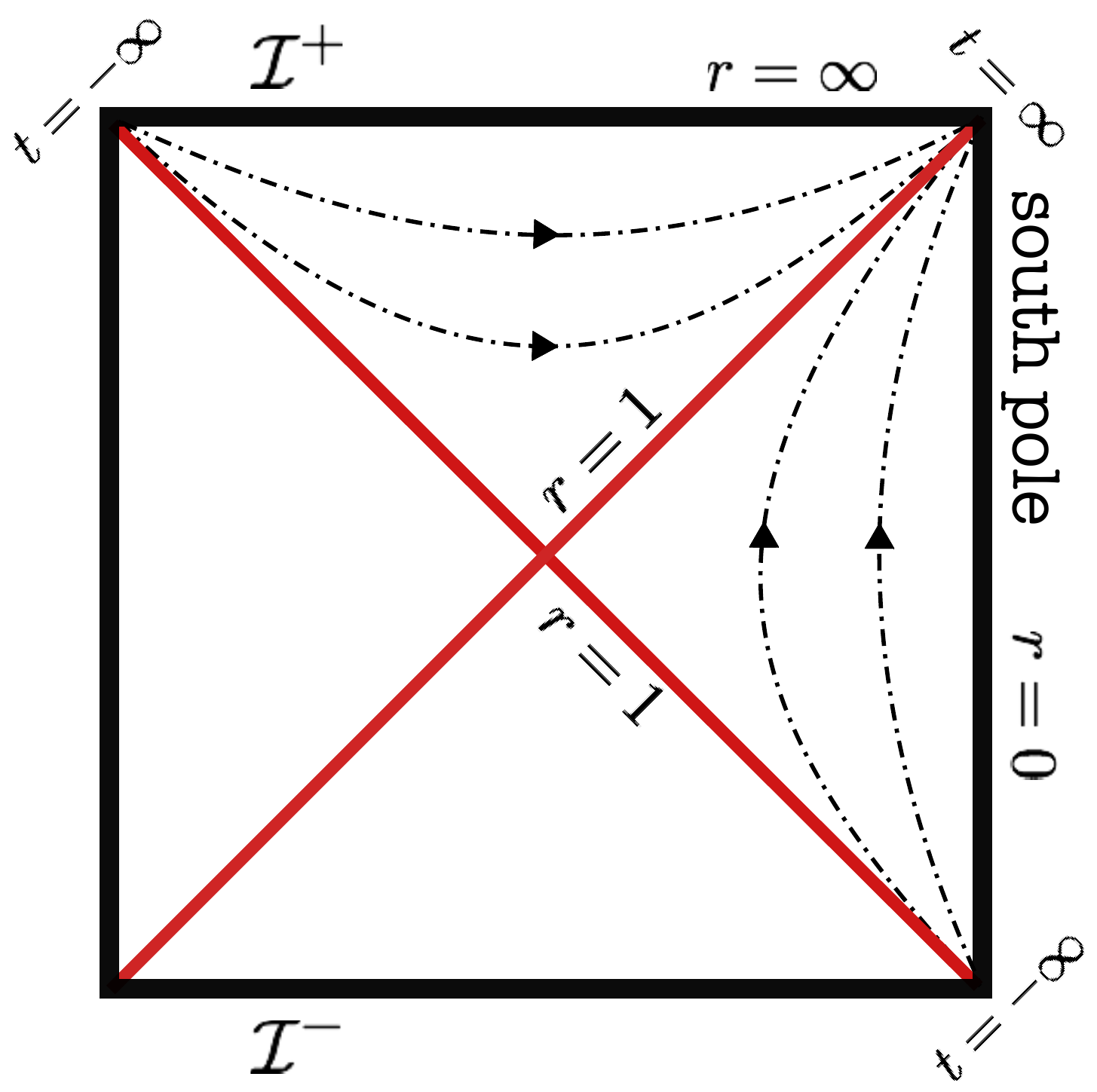}\hfill
\includegraphics[height=6.4cm]{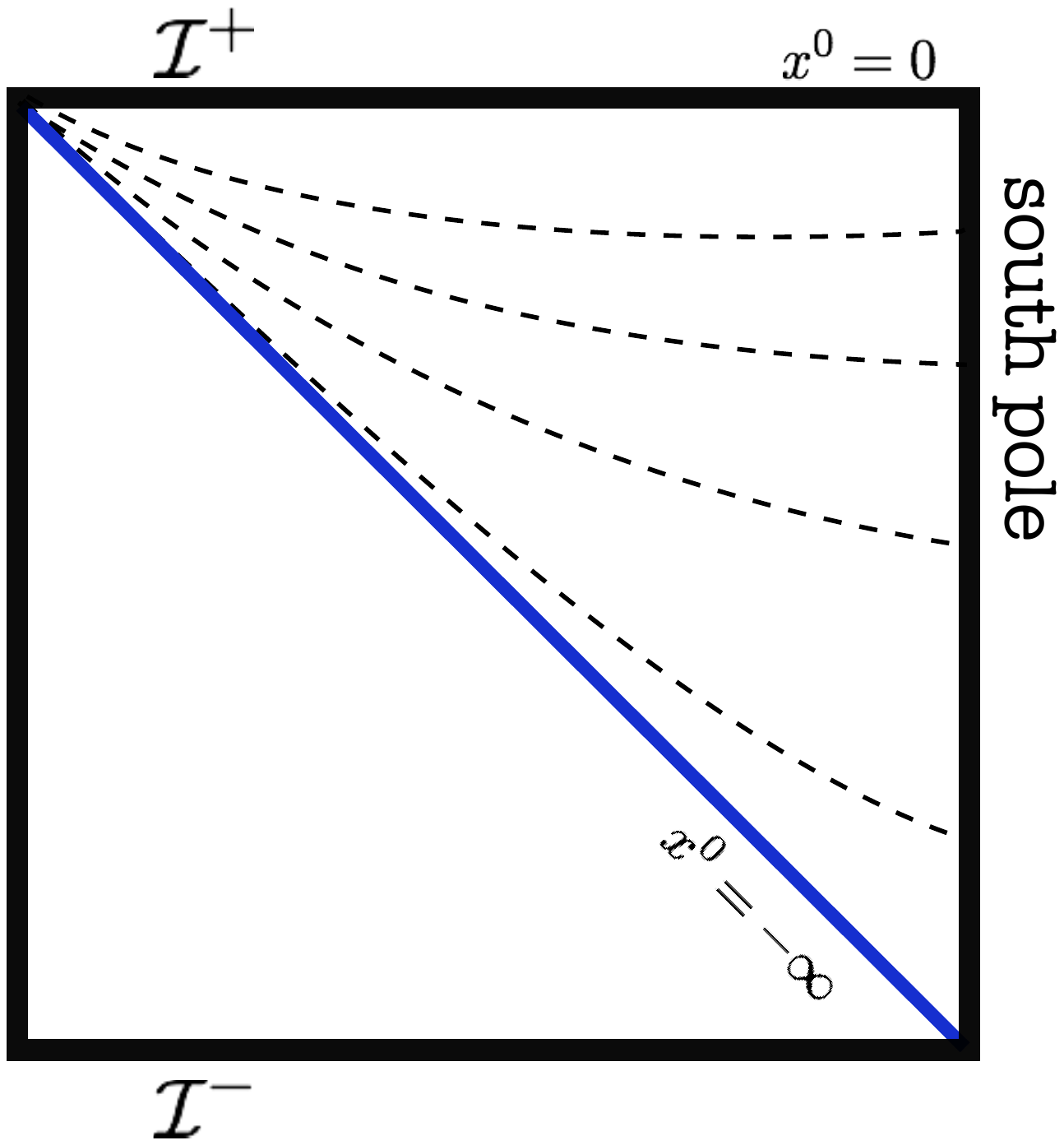}
}\caption{Here we present the de Sitter Penrose diagram and how various coordinate systems cover the manifold. \textbf{Left:} Static patch coordinates cover one quarter of de Sitter space and are good for describing static observers whose worldline coincides with the south pole at $r=0$. The dash-dotted lines represent lines of constant $r$ and the arrows represent the flow of the time coordinate $t$. Note that constant $r$ slices become spacelike in the northern diamond. \textbf{Right:} Poincar\'e coordinates cover half of de Sitter space. The dashed lines represent spacelike surfaces of constant conformal time $x^0$. }\label{Penrosediag}
\end{figure}
Because the static patch does not cover an entire Cauchy surface in de Sitter space, the QFT Hilbert space can be written as a tensor product ${\cal H} = {\cal H}_{in} \otimes {\cal H}_{out}$, where ${\cal H}_{in} ({\cal H}_{out})$ is the Hilbert space of degrees of freedom inside (outside) of the static patch.
The expectation value of an observable ${\cal O}$ which is inside the static patch can be written as ${\rm Tr}~{\rho {\cal O}}$, where $\rho$ is the reduced density matrix
\be\label{rhois}
\rho = \Tr_{{\cal H}_{out}}~ |0\rangle \langle 0|
\ee
obtained by tracing over the degrees of freedom outside the static patch. 
The standard Hawking radiation computation  \cite{ Gibbons:1977mu,Bousso:2001mw} then gives
\be\label{rho2}
\rho = {1\over Z} e^{-2 \pi a H}~~~~~~H = {\partial \over \partial t}
\ee
where $H$ is the Hamiltonian which generates time translation in the static time coordinate $t$. We emphasize that this Hawking radiation computation is correct, even though the vacuum state $|0\rangle$ is SUSY invariant.

Equation (\ref{rho2}) is the standard form for a finite temperature density matrix.  However, we note that the interpretation of (\ref{rho2}) is slightly different from that of a finite temperature density matrix in flat space.  In flat space, the matrix density (\ref{rho2}) would break SUSY if $H$ were the standard Hamiltonian which generates time-translations.  For example, correlation functions in this vacuum would not obey standard SUSY Ward identities.  A simple way to see this is to note that this Hamiltonian is the square of a supersymmetry generator, so its expectation value would vanish in an SUSY invariant state.  In de Sitter space, however, the operator $H$ which generates static-patch time translation is {\it not} the square of a supersymmetry generator.  Thus it is not subject to the usual requirement that it vanish in a supersymmetric state.  Indeed, none of the $SO(4,1)$ de Sitter isometries can be written as the square of a fermionic generator. The only operators which can be so written (and hence vanish in the ground state) are generators of conformal isometries in $SO(4,2)$ which are not regular de Sitter isometries.
For example, the generator $K_0$ of conformal time translation vanishes in the ground state.

An analogous situation would arise in flat space if we considered an observer undergoing constant acceleration.  In this case, the usual SUSY invariant flat space vacuum state $|0\rangle$ would appear to emit a finite temperature bath of Hawking quanta.
Indeed, observables in the causal region associated with the accelerating observer (the Rindler wedge) are computed precisely with the density matrix (\ref{rhois}-\ref{rho2}), where $H$ is the Hamiltonian which generates translations in the Rindler time coordinate. 
This finite temperature radiation is present even though the vacuum state exactly preserves supersymmetry.

\section{Acknowledgments}  We thank  Dionysios Anninos,  Bernard de Wit, Lorenzo di Pietro, 
Thomas Dumitrescu, Jaume Gomis, Daniel Jafferis,  Matthew Headrick and Andrew Strominger
for very useful discussions.  The research of AM is supported by NSERC, New Frontiers in Astronomy and Cosmology, the Foundational Questions Institute and the Simons Foundation. The research of DZF is supported in part by NSF grant PHY-0967299.  Both DZF and TA are supported in part by the U.S. Department of Energy under cooperative research agreement DE-FG02-05ER41360.


\appendix

\section{Spinor and Dirac matrix conventions}\label{conventions}
Throughout the main text we use the conventions of \cite{Freedman:2012zz}. Dirac matrices satisfy
\be
\lbrace \g^\m,\g^\n \rbrace=2g^{\m \n} ~.
\ee
We use greek indices $\m,\n\dots$ to denote curved space Dirac matrices, namely those which contain a frame field $\g^\m\equiv e^\m_a\g^a$.  Latin indices $a,b,\dots$ or hatted indices $\hat{\m},\hat{\n}\dots$ denote the local Lorentz frame. We define
\be
\g_{\m\n}\equiv\frac{1}{2}[\g_\m,\g_\n]~.
\ee
We use $\g_5\equiv i\g_{\hat 0}\g_{\hat 1}\g_{\hat 2}\g_{\hat 3}$ as well as the projection operators
\be
P_L=\frac{1+\g_5}{2}~,~~~~~~~~~~~~~~P_R=\frac{1-\g_5}{2}~.
\ee
Finally the Dirac adjoint of a fermionic field such as $\chi$ is given by
\be
\bar{\chi}\equiv i\chi^\dagger\g^{\hat{0}}~.
\ee

\section{$\cn=4$ SYM  with manifest $SU(4)_R$  symmetry}\lab{n4ap}

Radiative corrections in the $\cn =1$ theories discussed in the main text break conformal and superconformal symmetry.  However, the  $\cn=4$ theory  \cite{Brink:1976bc}, \cite{Gliozzi:1976qd} 
is an
exact superconformal theory in Minkowski spacetime and this property is expected to be maintained in dS$_4$.  Therefore we present the de Sitter extension of this theory in a manifestly $SU(4)_R$ invariant form closely related to the notation and  formulation in \cite{Gliozzi:1976qd}.

The theory contains the gauge potential $A_\m^a$, four Majorana gauginos whose chiral projections are $P_L\l^a_\a,~P_R\l^{a\a}$,  and six real scalars $X^i$.
We place the $SU(4)$ index down or up according to chirality.
The Lagrangian consists of the sum of a quadratic kinetic term, a cubic Yukawa term and quartic scalar potential: 
\be \lab{N4kinetic}
\cl_2 = -\bigg[ \frac14 F_{\m\n}^aF^{\m\n a} +  \bar{\l}^{a\a} \g^\m D_\m P_L \l^a_\a +\frac12 D_\m X^a_i D^\m X^a_i +\frac{1}{a^2} X^a_i X^a_i\bigg]\,.
\ee
\be \lab{cubic}
\cl_3 =-\frac12f^{abc}X^a_i[C_i^{\a\b} \bar{\l}^{b}_\a P_L\l^c_\b  + C_{i\a\b}  \bar{\l}^{b\a}P_R \l^{c\b}]\,.
\ee
\be \lab{quartic}
\cl_4 = - \frac14 f^{abc}f^{ab'c'} X^b_iX^c_j \,X^{b'}_iX^{c'}_j \,.
\ee

The cubic Lagrangian contains the (modified) set of six  't Hooft instanton matrices $C_i$.  The $C_i$ are real when $i=1,2,3$ and imaginary when $i=4,5,6$.  The $4\times 4 $ anti-symmetric matrices $C_i$ are
\begin{equation}\label{alphabeta}
\begin{array}{lll}
C_1 = 
\begin{pmatrix}
  0 & \sigma_1 \\ -\sigma_1 & 0 
\end{pmatrix}\,, &
C_2 = 
\begin{pmatrix}
   0 & -\sigma_3 \\ \sigma_3 & 0 
\end{pmatrix}\,, &
C_3 = 
\begin{pmatrix}
   i \sigma_2 & 0 \cr 0 &  i \sigma_2 
\end{pmatrix}\,, \\[4mm]
C_4 = -i
\begin{pmatrix}
   0 & i \sigma_2  \cr i \sigma_2  & 0 
\end{pmatrix}\,, &
C_5 = -i
\begin{pmatrix}
   0 & 1 \cr -1 & 0 
\end{pmatrix}\,, &
 C_6 =-i 
\begin{pmatrix}
   -i \sigma_2 & 0 \cr 0 &  i \sigma_2 
\end{pmatrix}\,,
\end{array}
\end{equation}
and $\sigma_i$ are the usual Pauli matrices.
The $C_i$ matrices are written as $C_i^{\a\b}$ when applied to left-handed spinors and as $C_{i\a\b} \equiv (C_i^{\a\b})^*$ when applied to right-handed spinors.   Note that the operator $[C_i^{\a\b} \bar{\l}^{b}_\a P_L\l^c_\b  + C_{i\a\b}  \bar{\l}^{b\a}P_R \l^{c\b}]$ is hermitian. 

The action is invariant under the transformation rules that involve an $SU(4)$ quartet of Majorana conformal Killing spinors $P_L \e_\a, ~ P_R\e^\a$:
\bea
\d A^a_\m &=& - \bar\e^\a\g_\m P_L\l^a_\a- \bar\e_\a\g_\m P_R\l^{a\a}\\
\d X^a_i &=&- [\bar\e_\a P_L C_i^{\a\b}\l_\b + \bar\e^a P_R C_{i \a\b} \l^{a\b}]\\
\d\l_\a^a &=&[ \frac{1}{2} \g^{\r\s}F_{\r\s}^a \e_\a -(\g^\m D_\m X_i)(P_L C_i^{\a\b}\e_\b + P_R C_{i \a\b}\e^\b)\\
&-&\frac12 X_i(P_R C_i^{\a\b}\slashed{D}\e_\b+  P_L C_{i \a\b}\slashed{D}\e_\b)] +  \D\l^a_\a\\
\D\l_\a^a &=& -\frac12 f^{abc}X^b_iX^c_j [(C_iC_j)^\a{}_\b P_R\e^\b + (C_iC_j)_\a{}^\b P_L\e_\b]\,.
\eea
The last term contains the effects of $F$ and $D$ aux. fields in the $\cn=1$ description.  To be clear on notation, we write
$(C_iC_j)^\a{}_\b= C^{\a\g}_iC_{j\g\b}$ and $(C_iC_j)_\a{}^\b= C_{i\a\g}C_{j}^{\g\b}$. \\
The SUSY parameters $\e$ in these formulas are scaled by a factor of $\sqrt2$ compared to those in the main text.  Note that the formulas for $\d\l$ and $\D\l$ contain both chiralities and do not fully conform to the down/up convention stated above.

\section{Conformal Killing spinors}\label{CKSAp}
We present here a compendium of formulae satisfied by conformal Killing spinors. Since we wish to be as general as possible, we given our presentation in arbitrary dimension $d$. 

Conformal Killing spinors are defined to live in the kernel of the Penrose operator 
\be
P_\m=D_\m-\frac{1}{d}\g_\m\slashed{D}~.
\ee
Note that any CKS $\e$ satisfying $P_\m\e=0$ also satisfies $D^\m P_\m\e=0$, and therefore
\be\lab{boxeps}
\Box\e=\frac{1}{d}\slashed{D}^2\e=\frac{1}{d}\left(\Box-\frac{1}{4}R\right)\e~,
\ee 
where the second equality in (\ref{boxeps}) can be derived using the integrability condition $[D_\m,D_\n]\e=\tfrac{1}{4}R_{\m\n a b}\g^{a b}\e$. We therefore determine that a CKS $\e$ must also satisfy additional constraints given by
\be\lab{eignsq}
\Box\e=-\frac{R}{4(d-1)}\e~,\quad \quad\slashed{D}^2\e=-\frac{d\,R}{4(d-1)}\e~.
\ee
A conformal Killing spinor is also a Killing spinor if it satisfies the added condition $\slashed{D}\e=\lambda\e$ for some constant $\lambda$. From (\ref{eignsq}) we determine
\be
\lambda_\pm=\pm\sqrt{-\frac{d\,R}{4(d-1)}}
\ee
which are real for any spacetime of constant negative curvature, and pure imaginary for spacetimes of constant positive curvature. This matches equations (\ref{adsksp}) and (\ref{dsksp}). Note that in a constant curvature spacetime, given a conformal Killing spinor $\e$, one can construct a pair of Killing spinors
\be\lab{Killingfromconf}
\e^{\pm}=\e \mp \sqrt{-\frac{4(d-1)}{d\,R}}\slashed{D}\e
\ee
with eigenvalues $\lambda_\pm$. 

Using the integrability conditions one can further derive 
\be\lab{schouten}
D_\m \slashed{D}\e=\frac{d}{2(d-2)}\left(-R_{\m\n}+\frac{1}{2(d-1)}g_{\m\n}R\right)\g^\n\,\e~.
\ee
 
\subsection{Maximally symmetric spacetimes}
In any maximally symmetric spacetime where $R_{\m\n}=\tfrac{R}{d}g_{\m\n}$, equation (\ref{schouten}) simplifies considerably
\be\lab{twoderiv}
D_\m \slashed{D}\e=-\frac{R}{4(d-1)}\g_\m\,\e=\frac{1}{d}\g_\m\slashed{D}^2\e~.
\ee
Thus, if $\e$ is a CKS for some maximally symmetric spacetime, then so is $\slashed{D}\e$. In lieu of (\ref{Killingfromconf}), this makes it clear that the real and imaginary parts of any Killing spinor in de Sitter space give two independent conformal Killing spinors. 

We use this to derive
\be
D_\m D_\n \e=-\frac{R}{4d(d-1)}\g_\n\g_\m\,\e
\ee
and hence $D_{(\m}D_{\n)}=-\tfrac{R}{4d(d-1)}g_{\m\n}\,\e$ and $D_{[\m}D_{\n]}=\tfrac{R}{4d(d-1)}\g_{\m\n}\,\e$. From (\ref{twoderiv}) we may also derive
\be
 \slashed{D}D_\m\e=\frac{R(d-2)}{4d(d-1)}\g_\m\,\e~,
\ee
and finally
\be
[D_\m,\slashed{D}]\e=-\frac{R}{2d}\g_\m \e~.
\ee


 \end{document}